\documentstyle[aps]{revtex}
\begin{document} 
\title{The inverse Mermin-Wagner theorem for classical spin models on
graphs}
\author{Raffaella Burioni\footnote{E-mail:
burioni@pr.infn.it}, Davide Cassi\footnote{E-mail: cassi@pr.infn.it},
Alessandro Vezzani\footnote{E-mail:vezzani@pr.infn.it}}
\address{Istituto Nazionale di Fisica della Materia, Dipartimento di
Fisica\\ 
Universit\`a di Parma, Parco Area delle Scienze n.7A, 43100 Parma, Italy  }
\date{\today}
\maketitle
\begin{abstract}
In this letter we present the inversion of the Mermin-Wagner theorem on graphs,
by proving the existence of spontaneous magnetization at finite temperature 
for classical spin models on transient on the average (TOA) graphs, i.e. 
graphs where a random walker returns to its starting point with an
average probability $\bar F < 1$. 
This result, which is here proven for models with $O(n)$ symmetry, includes
as a particular case $n=1$, providing a very general
condition for spontaneous symmetry breaking on inhomogeneous structures 
even for the Ising model.
\end{abstract}


Geometry plays a fundamental role in phase transitions of 
statistical models on regular lattices. The existence itself of an ordered
phase at non zero temperature only depends on large scale topology, via
the Euclidean dimension $d$ of the lattice. 
Indeed, a discrete symmetry is broken if and only if $d>1$, while 
for a continuous symmetry the corresponding condition is $d>2$.
For the latter case two rigorous results,  the Mermin-Wagner 
theorem \cite{mw1,mermin} and the Fr\"olich-Simon-Spencer bound
\cite{fss1,fss2}, provide respectively the necessary and the sufficient 
condition for spontaneous symmetry breaking.

This simple and exhaustive picture allows to classify a 
statistical system on a lattice in terms of 
{\it geometrical superuniversality classes} characterized by the 
Euclidean dimension. 

Unfortunately the classification cannot be 
directly extended to general networks describing non-crystalline structures,
where one cannot exploit the basic geometrical features of crystal lattices, 
such as translation invariance, the concept of Euclidean dimension, the 
reciprocal lattice. The proofs given in \cite{mw1,mermin,fss1,fss2} strongly 
rely on these properties and more general concepts and tools are needed when 
dealing with a non-crystalline structure.

A basic improvement in the study of properties of geometrically disordered 
structures has been achieved with graph theory. 
A graph, i.e. a general network of sites connected pairwise by bonds,  
provides the most suitable mathematical tool to describe complex and irregular 
discrete geometries. Euclidean lattices, which are the usual model for 
crystalline structures, are very peculiar example of graphs
characterized by complete translation invariance.

The generalization to inhomogeneous structures of the necessary condition
for spontaneous breaking of continuous symmetry, the 
Mermin-Wagner theorem, has been the first step in this 
direction \cite{mwg,mwg2}. 
The existence of spontaneous magnetization on a graph ${\cal G}$ is 
related to the probability $F_i$ of returning to the starting site $i$ for a 
simple random walk on ${\cal G}$. 
In particular, it was proven that there is no spontaneous magnetization 
for {\it recurrent on the average} (ROA) graphs; i.e. when $\bar{F}=1$,
where $\bar F$ is the 
average of $F_i$ over all the points $i$ of the graph ${\cal G}$ \cite{mwg}. 
This result naturally includes the lattice theorem \cite{mw1}, since Euclidean 
lattices in 1 and 2 dimensions turn out to be ROA. However up to now a 
sufficient condition has been lacking.

In this letter we study the case $\bar{F}<1$, i.e. 
{\it transient on the average} (TOA) graphs and we give a rigorous proof 
of the existence of spontaneous magnetization at $T>0$ for classical spin 
models with $O(n)$ symmetry. This result is the exact inversion
of theorem \cite{mwg} for the classical case and a generalization to 
graphs of \cite{fss1,fss2}, since lattices with $d>2$ are TOA. Now, each graph 
can be classified either as ROA or TOA and therefore this theorem completes the
picture for classical spin models on graphs.
Moreover, as in the lattice case \cite{fss1,fss2}, the proof also holds  
for $n=1$, i.e. for the Ising model.   

In the following ${G}$ is a graph consisting of $N_g$ sites, 
$i=1,2,...N$, and of links $(ij)$ joining them; we say that two sites  
connected by a bond are nearest neighbors. A graph is connected if, given
any two points in ${G}$, there exists a path joining them.
Here we will consider connected graphs. The chemical distance between 
sites $i$ and $j$ is the length (number of links) of the shortest path 
joining them. The graph topology is algebraically
described by its adjacency matrix $A_{ij}$, given by:
$A_{ij}= 1 $ if  $i$ and $j$ are nearest neighbors, $A_{ij}= 0$ otherwise.
$O(n)$ models on ${\cal G}$ with  $n \geq 1$ are defined by the
Hamiltonian:
\begin{equation}
{\cal H} = - {1\over 2}\sum_{ij} J_{ij}
{\overrightarrow{\sigma}}_i 
\cdot {\overrightarrow{\sigma}}_j  -\overrightarrow{h} \cdot \sum_{i} 
{\overrightarrow{\sigma}}_i
\label{defH}
\end{equation}
where $J_{ij}$ are bounded ferromagnetic interactions on ${G}$:
\begin{equation}
J_{ij}=J_{ji} = \left\{
\begin{array}{cl}
J_{ij} & {\rm with}  \ 0<\epsilon\leq J_{ij}\leq J<\infty \ {\rm  if\ } 
A_{ij}=1 \cr
0      &  \ {\rm  if\ } A_{ij}=0 \cr 
\end{array}
\right .
\end{equation}
and $z_i=\sum_j J_{ij} \leq z < \infty.$
${\overrightarrow{\sigma}}_j$ are $n$-dimensional real unit vectors 
${\overrightarrow{\sigma}}_i \equiv (\sigma^1,\ldots,\sigma^n)$ 
defined on each vertex satisfying the constraints:
$|{\overrightarrow{\sigma}}_i|^2=1 ~~ \forall i $.
For $n=1$ ${\cal H}$ describes the Ising model which is invariant under
the discrete 
symmetry group ${\bf Z_2}$, while for $n \geq 2$ ${\cal H}$ represents a
model with an $O(n)$ continuous symmetry. 
Finally $\overrightarrow{h} \equiv (h,0,\ldots,0)$, $h > 0$, is an 
external magnetic field coupled to $\overrightarrow{\sigma}_i$.

The average magnetization is the order parameter of the model and it is
defined by:
\begin{equation}
M(h) \equiv {1\over N_g} \sum_i \langle \sigma_i^1\rangle.
\label{defM}
\end{equation}
where the average $\langle \rangle$ is taken with respect to 
the usual Boltzmann weight $\exp(-\beta {\cal H})$ with $\beta=1/KT$. 

The Mermin-Wagner theorem will be inverted by proving a lower positive bound 
for the magnetization at sufficiently low $T$ in the thermodynamic 
limit $N_g\rightarrow \infty$, on an infinite graphs $\cal{G}$. Namely, we will 
show that if $\bar{F}<1$, it exists a small enough temperature $T$ 
for which $\lim_{h\rightarrow 0}M(h)> c(T)$ with $c(T)>0$.
In order to define the thermodynamic limit let us introduce 
the Van Hove spheres $S_{r,o}$ as the subsets of sites in $\cal{G}$, whose 
chemical distance from $o$ is $\le r$, $N$ is the number of sites in 
$S_{r,o}$. In our proof to explore the behavior of the model on the infinite 
graph, first we will obtain inequalities for thermal averages on the finite  
subgraphs ${\cal{S}}_{r,o}$ given by the sites of the sphere $S_{r,o}$ and 
the bounds $(i,j)$ of $\cal{G}$ with $i,j,\in S_{r,o}$. Then we will take the 
thermodynamic limit $N_g\rightarrow \infty$, letting $r\rightarrow \infty$; 
finally we will take the limit $h \rightarrow 0$.

Let us first consider graphs for which the average of $F_i$ is smaller than 
$1$ in every positive measure subset ${S}$ of the sites of $\cal{G}$, 
where the measure of the subset ${S}$ is given by 
$|{S}|=\lim_{N_g \rightarrow \infty}[\sum_{i} {\chi_{{S}}}(i)]/N_g$ and
${\chi_{{S}}(i)}$ is the characteristic function of ${S}$:
${\chi_{{S}}(i)}=1$ if $i\in {S}$ and ${\chi_{{S}}(i)}=0$ if 
$i \not\in {S}$. We will 
call these graphs pure TOA. For these graphs  
in the thermodynamic limit $N_g\rightarrow \infty$ \cite{bcv}:
\begin{equation}
\lim_{\mu\rightarrow 0} \lim_{N_g\rightarrow \infty} {1 \over N_g} 
{\rm Tr} (L+\mu)^{-1}=v<\infty. 
\label{trace}
\end{equation}
where $L_{ij}$ is the Laplacian operator given by 
$L_{ij}=z_i\delta_{ij} - J_{ij}$ and $\mu_{ij}= \mu \delta_{ij}$, $\mu >0$.

Our proof will follow these main steps:

a) we introduce for the constraints $|{\overrightarrow{\sigma}}_i|^2=1$ 
an integral representation with Lagrange multipliers $\alpha_i$ and 
perform the Gaussian integration on $\overrightarrow{\sigma}_i$.

b) we determine the asymptotic behavior of the integrals over $\alpha_i$ 
for $\beta \rightarrow \infty$ by saddle point technique.

c) we establish the lower bound on $M(h)$ exploiting b) and
the identity:
\begin{equation}
1 = {1\over N_g} \sum_i \langle |\overrightarrow{\sigma}_i|^2 \rangle.
\label{glcon}
\end{equation}

Let us start with step a).
In the expressions (\ref{defM}) and (\ref{glcon}) we introduce the
integral representation for the constraints 
$|{\overrightarrow{\sigma}}|_i^2=1$:
\begin{equation}
\delta(|\overrightarrow{\sigma}_i|^2-1)={e^{\epsilon/2} \over 2 \pi}
\int d\alpha_i~e^{[-i\alpha_i (|\overrightarrow{\sigma}_i|^2-1)/2
-\epsilon |\overrightarrow{\sigma}_i|^2/2]}
\end{equation}
where $\epsilon$ is a real arbitrary constant. We will chose 
$\epsilon=h\beta$. 
We now perform the Gaussian integration over the variables 
$\overrightarrow{\sigma}_i$, obtaining for (\ref{defM}) and 
(\ref{glcon}):
\begin{equation}                                    
M(h) ={1\over Z}{\int \prod_{i\in S_{r,o}} d \alpha_i \ 
e^{i S_{\beta h}(\alpha)}
\ {h\over N_g} \sum_{kj}(L+H+i\alpha)^{-1}_{kj}}
\label{defm1}
\end{equation}

\begin{equation}
1 = {1\over Z} \int \prod_{i\in S_{r,o}} d \alpha_i \ e^{i S_{\beta
h}(\alpha)} [ 
{n \over \beta N_g} {\rm\ Tr}(L+H+i\alpha)^{-1} +
{h^2\over N_g} \sum_{ij}(L+H+i\alpha)^{-2}_{ij} ].
\label{glconst1}
\end{equation}
where:
$$ i S_{\beta h}(\alpha)\equiv
-{n\over 2} {\rm\ Tr} (\ln (L+H+i\alpha)) 
+{\beta\over 2}
[i\sum_i\alpha_i+h^2\sum_{ij}(L+H+i\alpha)^{-1}_{ij}],$$ 
$$Z\equiv \int \prod_{i\in S_{r,o}} d \alpha_i \ e^{i S_{\beta
h}(\alpha)},
\ \ \alpha_{ij}=\alpha_i\delta_{ij} \ \ \ {\rm and} 
\ \ \ H_{ij}=h\delta_{ij}.$$
Notice that the order of the symmetry group $n$ becomes a parameter of
the integration.

Let us now study the behavior of (\ref{defm1}) and (\ref{glconst1})
for large $\beta$, that is point b) of our plan. By saddle point
theorem,
the leading asymptotic behavior of (\ref{defm1}) and (\ref{glconst1}) is
given by the $\bar {\alpha_i}$ which satisfy the stationary conditions:
\begin{equation}        
{\partial \over \partial \bar{\alpha}_i}
[i\sum_k\bar{\alpha}_k+h^2\sum_{kj}(L+H+i\bar{\alpha})^{-1}_{kj}]=0 
\ \ \forall\ i
\label{statcon}
\end{equation}
where $\bar{\alpha}_{ij}=\bar{\alpha}_i\delta_{ij}$.
Equation (\ref{statcon}) is satisfied for all $h \geq 0$
only if $\bar{\alpha}_i=0\ \forall\ i$, so that \cite{bcv}:
\begin{equation}        
M(h) ={1\over Z}\int_{\Gamma} \prod_{i\in S_{r,o}} d \alpha_i 
\ {\rm Re} [e^{i S_{\beta h}(\alpha)}]
\ {h\over N_g} {\rm Re} [\sum_{kj}(L+H+i\alpha)^{-1}_{kj}] + o(1/\beta)
\label{defm2}
\end{equation}
\begin{equation}
1 = {1\over Z} \int_{\Gamma}  \prod_{i\in S_{r,o}} d \alpha_i 
\ {\rm Re} [e^{i S_{\beta h}(\alpha)}] {\rm Re} [ 
{n \over \beta N_g} {\rm\ Tr}(L+H+i\alpha)^{-1} +
{h^2\over N_g} \sum_{ij}(L+H+i\alpha)^{-2}_{ij}]+ o(1/\beta).
\label{glconst2}
\end{equation}
where $\Gamma$ is the region around the saddle point $\bar{\alpha_i}$ in
which
${\rm Re} (\exp(i S_{\beta h}(\alpha))) > 0$. Here we exploited the
property
that $\exp(i S_{\beta h}(\bar{\alpha_i}))$ is real and positive and 
that we are evaluating real quantities.

As for c), we first introduce the following inequalities, which can be
proven exploiting the boundedness and the non negativity of the Laplacian 
operator \cite{bcv}: 
\begin{equation}
1 \geq {\rm Re} [ {h\over N_g} \sum_{ij}(L+H+i\alpha)_{ij}^{-1} ]\geq
{\rm Re} [{h^2\over N_g} \sum_{ij}(L+H+i\alpha)_{ij}^{-2}] 
\label{ineq1}
\end{equation}
\begin{equation}
0 \leq {1\over N_g}{\rm Re}[{\rm Tr}(L+H+i\alpha)^{-1}] \leq
 {1\over N_g}{\rm Tr}(L+H)^{-1}.
\label{ineq2}
\end{equation}
Using (\ref{ineq1}) we compare  the expressions (\ref{defm2}) and 
(\ref{glconst2}), obtaining for the magnetization: 
\begin{equation}
M (h)\geq 1 - o(1/\beta)-  
{1\over Z} \int_{\Gamma}  \prod_{i\in S_{r,o}} d \alpha_i\ {\rm Re}
[e^{i S_{\beta h}(\alpha)}] 
{n \over \beta N_g} {\rm Re}[{\rm Tr}(L+H+i\alpha)^{-1}].      
\label{M3}
\end{equation}
Now with (\ref{ineq2}) we obtain for $M(h)$ the following inequality:
\begin{equation}
M (h)\geq 1 - o(1/\beta)-  
{1\over N_g}{\rm Tr}(L+H)^{-1}.     
\label{M4}
\end{equation}
Using property (\ref{trace}) of pure TOA graphs, we finally get in the 
thermodynamic limit:
\begin{equation}
\lim_{h\rightarrow 0} \lim_{N_g\rightarrow \infty}M (h) \geq 
1 -  {v \over \beta} - o(1/\beta)  
\label{M5}
\end{equation}
and this complete the proof for pure TOA graphs.

Let us consider now the most general case of a graph $\cal{G}$ which is not 
pure TOA. In this case $\cal{G}$ must have a positive measure subset
where the average of $F_i$ is 1, i.e a ROA  
subgraph. We call such a graph mixed TOA. A mixed TOA graph can always be 
decomposed in a pure 
subgraph $\cal{S}$ and a ROA subgraph, connected by a zero measure set of 
links. This implies that the total free energy per site $f_{\cal{G}}$
is given by: 
$f_{{\cal{G}}}=|{\cal{S}}|f_{\cal{S}}+|{\cal{G}}-
{{\cal{S}}}|{f}_{{\cal{G}}-{\cal{S}}}$ and as a consequence:
\begin{equation}
\lim_{N_g \rightarrow \infty}M(h)\geq \lim_{N_g \rightarrow \infty} 
{1\over N_g} \sum_{i\in \cal{S}} \langle \sigma_i^1 \rangle
\label{mags}
\end{equation}
On   $\cal{S}$ :
\begin{equation}
\lim_{h\rightarrow 0}{1 \over N_g} \sum_{i\in
\cal{S}}(L+H)^{-1}_{ii}=v'<\infty. 
\label{trace2}
\end{equation}
and using (\ref{M5}), (\ref{mags}) and (\ref{trace2}) we get:
\begin{equation}
\lim_{h\rightarrow 0}\lim_{N_g \rightarrow \infty} M (h)  
\geq |{\cal{S}}| - {v'\over \beta}-o(1/\beta). 
\label{M6}
\end{equation}
Inequality (\ref{M6}) proves the existence of a lower positive bound at 
low enough temperature for the magnetization of an $O(n)$ model defined
on a TOA graph. In this way we obtain the inversion of \cite{mwg} and 
we generalize the Fr\"olich-Simon-Spencer result to generic
inhomogeneous discrete structures.

A few comments follow from our result.
The condition ${\bar F} < 1$ turns out to be a condition on
the spectral density at low eigenvalues of the Laplacian operator $L$ on 
${\cal G}$ and provides the link between 
the physical properties of the $O(n)$ model and the topology of the 
discrete space. In particular it includes the lattice {\it geometrical
superuniversality class} $d > 2$, i.e the result of \cite{fss1,fss2}.
More generally for ROA and pure TOA graphs, if  one can define the 
spectral dimension $\bar d$ \cite{univ}, the condition
becomes ${\bar d} > 2$. However we point out that the present result is far
more general, holding also for graphs without spectral dimension. This is
the case of the Bethe lattice, which is a pure TOA graph with finite 
temperature phase transitions.

Our result completes the description of the behavior of 
continuous classical spin models on generic networks. On the other hand 
it also provides a rigorous and very general sufficient condition  
for spontaneous magnetization of the Ising model ($n=1$) on graphs. 
Obviously this condition is not necessary. A simple counterexample 
is the two dimensional Ising model, which has spontaneous magnetization. 
The study of the Ising model on ROA graphs is therefore a key step  
to obtain a complete picture of the behavior 
of spin models on general discrete structures.



\begin{thebibliography}{99}
\bibitem{mw1}
N.D. Mermin and H. Wagner, {\it Phys. Rev} {\bf 17} (1966) 1133
\bibitem{mermin}
N.D. Mermin, {\it Journ. of Math. Phys.} {\bf 8} 1061 (1967) 
\bibitem{fss1}
J. Froelich, B. Simon and T. Spencer, {\it Phys. Rev. Lett.} {\bf 36}
(1976) 804
\bibitem{fss2}
J. Froelich, B. Simon and T. Spencer, {\it Commun. Math. Phys.} {\bf 50} 
(1976) 79
\bibitem{mwg}
D. Cassi, {\it Phys. Rev. Lett.} {\bf 68} (1992) 3631; {\it Phys. Rev.
Lett.} {\bf 76} (1996) 2941
\bibitem{mwg2}
F. Merkl and H. Wagner, {\it J. Stat. Phys.} {\bf 75} (1994) 153. 
\bibitem{bcv}
R. Burioni, D.Cassi, A. Vezzani, preprint {\it cond-mat/9808046}
\bibitem{univ}
R. Burioni, D. Cassi {\it Phys. Rev. Lett.} {\bf 76} (1996), 1091;
{\it Mod. Phys. Lett.} {\bf B 11} (1997), 1095

\end{thebibliography}
\end{document}